\documentclass[aps,prd,nofootinbib,showkeys,showpacs,floatfix]{revtex4}
\usepackage{amsmath}
\usepackage{amstext}
\usepackage{amssymb}
\usepackage{graphicx}
\usepackage{rotating}
\usepackage{color} 
\usepackage{multirow} 
\usepackage[T1]{fontenc} 
\usepackage{slashed}

\def\gsim{\raise0.3ex\hbox{$\;>$\kern-0.75em\raise-1.1ex\hbox{$\sim\;$}}}
\def\lsim{\raise0.3ex\hbox{$\;<$\kern-0.75em\raise-1.1ex\hbox{$\sim\;$}}}

\newcommand{\eVq}  {\text{eV}^2}

\newcommand{\AddrAHEP}{
  {\it AHEP Group, Instituto de F\'{\i}sica Corpuscular --
    C.S.I.C./Universitat de Val{\`e}ncia \\
    Edificio de Institutos de Paterna, Apartado 22085,
  E--46071 Val{\`e}ncia, Spain}}

\begin{document}

\begin{flushright}
IFIC/12-31
\end{flushright}


\title{Global status of neutrino oscillation parameters after Neutrino-2012 }

\author{D.~V.~Forero} \email{dvanegas@ific.uv.es} \affiliation{\AddrAHEP}
\author{M. T{\'o}rtola}\email{mariam@ific.uv.es}    \affiliation{\AddrAHEP}
\author{J.~W.~F.~Valle} \email{valle@ific.uv.es} \affiliation{\AddrAHEP}
\begin{abstract}

  Here we update the global fit of neutrino oscillations in
  Refs.~\cite{Schwetz:2011qt,Schwetz:2011zk} including the recent
  measurements of reactor antineutrino disappearance reported by the
  Double Chooz, Daya Bay and RENO experiments, together with latest
  MINOS and T2K appearance and disappearance results, as presented at
  the Neutrino-2012 conference.
  We find that the preferred global fit value of $\theta_{13}$
    is quite large: $\sin^2\theta_{13} \simeq 0.025$ for normal
    and inverted neutrino mass ordering, with $\theta_{13} = 0$ now
    excluded at more than 10$\sigma$.
  The impact of the new $\theta_{13}$ measurements over the
    other neutrino oscillation parameters is discussed as well as the
    role of the new long-baseline neutrino data and the atmospheric
    neutrino analysis in the determination of a non-maximal
    atmospheric angle $\theta_{23}$.

\end{abstract}
\maketitle
\section{Introduction}

Recent measurements of $\theta_{13}$ have been reported by the reactor
experiments Double Chooz~\cite{Abe:2011fz}, Daya Bay~\cite{An:2012eh}
and RENO~\cite{Ahn:2012nd}.
These experiments look for the disappearance of reactor antineutrinos
over baselines of the order of 1~km, due to neutrino oscillations
mainly driven by the third mixing angle $\theta_{13}$ of the lepton
mixing matrix~\cite{schechter:1980gr,nakamura2010review}.
Up to now the most sensitive measurements of reactor antineutrinos
were reported by the past reactor experiments
CHOOZ~\cite{Apollonio:2002gd} and Palo Verde~\cite{Boehm:2001ik}.

Compared to their predecessors, the new reactor experiments have
larger statistics, thanks to their increased reactor power and the
bigger antineutrino detector size. On the other hand, one of their
most important features is that they have detectors located at
different distances from the reactor core. As a result measurements at
the closest detectors can be used in order to predict the expected
event number at the more distant detectors, avoiding the need to rely
on theoretical calculations of the produced antineutrino flux at the
reactors.
As a consequence these experiments have for the first time observed
the disappearance of reactor antineutrinos over short distances,
providing the first measurement of the mixing angle $\theta_{13}$, so
far unknown.

Last year there were some indications for a non-zero $\theta_{13}$
mixing angle coming from the observation of electron neutrino
appearance on a muon neutrino beam at the accelerator oscillation
experiments T2K~\cite{Abe:2011sj} and
MINOS~\cite{Adamson:2011qu}. Together with the hints from the solar
and atmospheric neutrino data samples, the global analysis of neutrino
oscillation data reported indications of non-zero $\theta_{13}$
between 3 and 4$\sigma$, depending on the treatment of short-baseline
reactor data in the full analysis (see Refs
\cite{Schwetz:2011qt,Schwetz:2011zk} for more details\footnote{
  Other recent global analyses previous to the Neutrino-2012
  conference can be found in Refs.\cite{GonzalezGarcia:2010er,Fogli:2011qn}}).
Here we update the global fit of neutrino oscillations given in
Refs.~\cite{Schwetz:2011qt,Schwetz:2011zk} by including the recent
measurements of reactor antineutrino disappearance reported by the
Double Chooz, Daya Bay and RENO experiments.
In addition to the above-mentioned data our analysis includes also
the most recent reactor neutrino data reported at the Neutrino-2012
Conference~\cite{new-DChooz,new-DayaBay}, as well as the latest
results of the MINOS~\cite{new-MINOS} and
T2K~\cite{Abe:2012gx,new-T2K-app} experiments.
The role of the new $\theta_{13}$ measurements upon the
determination of the remaining oscillation parameters is also
analyzed.
Particularly important is the impact of the new accelerator
neutrino data as well as the details of the atmospheric neutrino
analysis upon the determination of the atmospheric mixing angle
$\theta_{23}$.  

\vspace{3mm}

\section{Reactor experiments: Double Chooz, Daya Bay and RENO}
\label{sec:react-exper-double}
\label{sec:reac}

\vspace{2mm} 

This year, in chronological order, the Double Chooz, Daya Bay and RENO
Collaborations have reported measurements of the electron antineutrino
disappearance with important levels of statistical significance.

The Double Chooz (DC) experiment, located in France, is a reactor
experiment planned to have two detectors and two reactors. In its
first stage DC has reported 101 days of running~\cite{Abe:2011fz},
with only the far detector operating so far. The near detector (ND) is
expected to start operation by early 2013. The two reactors are
approximately equal, with an individual power of 4.25 GW$_\text{th}$
and are placed at a distance of 1050 m from the far detector. The
detector has a fiducial volume of 10 m$^3$ of neutrino target liquid.
From the analysis of the rate and the energy spectrum of the prompt
positrons produced by the reactor antineutrinos, the DC collaboration
find $\sin^2 2 \theta_{13} = 0.086 \pm 0.041 (\text{stat}) \pm 0.030
(\text{syst})$. Using only the ratio of observed to expected events
they get a slightly higher best fit value: $\sin^2 2 \theta_{13} =
0.104 \pm 0.030 (\text{stat}) \pm 0.076 (\text{syst})$.
A more recent analysis of DC data with an exposure of 227.93 live
days~\cite{new-DChooz} has reported the observation of 8249 candidate
electron antineutrinos while 8937 were expected in the absence of
oscillations. Using a rate plus spectral shape analysis the following
best fit value for the reactor angle is obtained: $\sin^2 2
\theta_{13} = 0.109 \pm 0.030 (\text{stat}) \pm 0.025 (\text{syst})$.

The Daya Bay (DYB) reactor experiment~\cite{An:2012eh} is a
neutrino oscillation experiment designed to measure the
mixing angle $\theta_{13}$ as well. 
The experiment is placed in China and it contains an array of three
groups of detectors and three groups of two-reactor cores. The far
group of detectors (far hall) is composed of three detectors and the
two near halls are composed by one and two detectors, respectively. In
order to reduce systematic errors, the detectors are approximately
equal, with a volume of 20 tons of Gadolinium-doped liquid scintillator
as neutrino target material.
The reactor cores are approximately equal as well, with a maximum
power of 2.9 GW$_\text{th}$ (total power of 17.4 GW$_\text{th}$) and the
distances to the detectors range from 350 to 2000 m
approximately.  The rate-only analysis performed by the DYB
collaboration finds a best fit value of $\sin^22\theta_{13}=0.092 \pm
0.016({\rm stat}) \pm 0.005({\rm syst})$. A zero value for
$\theta_{13}$ is excluded with a significance of $5.2\sigma$.
New results presented in the Neutrino-2012
Conference~\cite{new-DayaBay} with 2.5 times more statistics allow a
stronger rejection for $\theta_{13}$ = 0 that now is excluded at
almost 8$\sigma$ by DYB alone.  A rate-only statistical analysis
of the new DYB data reports a best fit value of
$\sin^22\theta_{13}=0.089 \pm 0.010({\rm stat}) \pm 0.005({\rm
  syst})$.

The RENO experiment~\cite{Ahn:2012nd} is situated in South Korea and
it has been running for 229 days. It shares some features with DC and
DYB. RENO has six reactor cores, distributed along a 1.3 km straight
line. Two of the reactors have a maximum power of 2.66 GW$_\text{th}$
while the other four may reach 2.8 GW$_\text{th}$. Reactor
antineutrinos are detected by two identical detectors, labeled as near
and far, located at 294 and 1383 m from the reactor array center. Each
RENO detector contains 16 tons of Gadolinium-doped Liquid
Scintillator. Based on a rate-only analysis, the RENO Collaboration
finds $\sin^2 2 \theta_{13} = 0.113 \pm 0.013({\rm stat.}) \pm
0.019({\rm syst.})$, together with a $4.9\sigma$ exclusion for
$\theta_{13} = 0$.

\vspace{3mm} 

 \section*{Reactor event calculation}
\label{sec:event-calculation}

Reactor antineutrinos are produced by the fission of the isotopes
$^{235}$U, $^{239}$Pu, $^{241}$Pu and $^{238}$U. Each fissile isotope 
contributes to the total reactor neutrino flux and fuel content with a
certain fission fraction $f^l$ that can be calculated through a 
detailed simulation of the core evolution. 
After their production, the reactor antineutrinos are detected at the
experiments via the inverse beta decay process, looking for a delayed
coincidence between the positron annihilation and the neutron capture
in the target material.  The window of positron energy covered by the
three experiments described above ranges from 0.7 to 12 MeV
approximately.

\vspace{2mm} 

For a given experiment, the total number of events expected at the $i$th
detector coming from the $r$th reactor can be calculated as:
\begin{equation}\label{event}
N_{i,r}=\frac{N_p\,P^r_{th}}{4\pi L_{ir}^2\,\langle E_{fis} \rangle}\epsilon_i\int_0^{\infty} dE_\nu \; \Phi^r(E_\nu) \sigma_{\text{IBD}}(E_\nu) \,P(E_\nu,L_i) 
\end{equation}
where $N_p$ is the number of protons in the target volume, $P^r_{th}$
is the total reactor power, $\epsilon_i$ denotes the efficiency of the
detector and $\langle E_{fis} \rangle = \sum_l f^l E^l_{fis}$ is the
average energy released per fission, calculated from the individual
fission fractions $f^l$ and the energy release per fission for a given
isotope $l$ taken from Ref.~\cite{Kopeikin:2004cn}.
For the antineutrino flux prediction $\Phi^r(E_\nu)$ we use the recent
parameterization given in Ref.~\cite{Mueller:2011nm} as well as the
new normalization for reactor antineutrino fluxes updated in
Ref.~\cite{Abazajian:2012ys}.
The inverse beta decay cross section $\sigma_{\text{IBD}}(E_\nu)$ is taken from
Ref.~\cite{Vogel:1999zy}. 
Finally, for the neutrino propagation factor $P(E_\nu,L_{ir})$ we
use the full three-neutrino disappearance probability.
The distance between reactor and detector $L_{ir}$ is also used to
correct the total antineutrino flux at the detector site.
In order to minimize the dependence upon the predicted normalization
of the antineutrino spectrum, we analyze the total rate of expected
events at the far detector/s in the presence of oscillations over the
no-oscillation prediction. This way, our statistical analysis is free
of correlations among the different reactor data samples, since the
relative measurements do not rely on flux predictions.

\section{Global analysis}
\label{sec:glob}

In our global analysis of neutrino oscillation parameters we combine
the recent reactor data from Double Chooz, Daya Bay and RENO with all
the remaining relevant experiments, as follows.

\subsection{Solar neutrino and KamLAND data}
\label{sec:solar-KamLAND}

We include the most recent solar neutrino data from the radiochemical
experiments Homestake~\cite{Cleveland:1998nv},
Gallex/GNO~\cite{Kaether:2010ag} and SAGE~\cite{Abdurashitov:2009tn},
as well as the latest published data from
Borexino~\cite{Bellini:2011rx}, the three phases of the
Super-Kamiokande
experiment~\cite{hosaka:2005um,Cravens:2008aa,Abe:2010hy} and the
three phases from the Sudbury Neutrino Experiment
SNO~\cite{aharmim:2008kc,Aharmim:2009gd}.
For our simulation of the production and propagation of neutrinos in
the Sun we consider the most recent update of the standard solar
model~\cite{Serenelli:2009yc}, fixing our calculations to the low
metallicity model labeled as AGSS09. The impact of the choice of a
particular solar model over the neutrino oscillation analysis has been
discussed in the arXiv updated version of Ref.~\cite{schwetz:2008er}.
We also include the most recent results published by the KamLAND
reactor experiment with a total livetime of 2135 days, including the
data collected during the radiopurity upgrade in the
detector~\cite{Gando:2010aa}.

\subsection{Atmospheric and accelerator neutrino data}
\label{sec:accel-exper}

In our global fit we use the atmospheric neutrino analysis done by the
Super-Kamiokande Collaboration~\cite{Wendell:2010md}. The oscillation
analysis has been performed within the one-mass scale approximation,
neglecting the effect of the solar mass splitting and  includes the
atmospheric results from the three phases of the Super-Kamiokande
experiment.
Concerning the long-baseline data, we include the most recent results
from the MINOS and T2K long-baseline experiments released last June at
the Neutrino 2012 Conference. We consider the appearance and
disappearance channels for both experiments as well as the neutrino
and antineutrino data for MINOS.
These new long-baseline results imply some improvements with respect
to the previous MINOS and T2K data in
Refs. \cite{Adamson:2011ig,Adamson:2011fa,Adamson:2012rm,Adamson:2011qu,Abe:2011sj}.
On the one hand, the new results on $\nu_\mu \to \nu_e$ appearance
searches allow a better determination of the $\theta_{13}$ mixing
angle, although its current determination is fully dominated by the
Daya Bay reactor data.
On the other hand, and here lies the most relevant implication of the
new long-baseline data, they show a preference for a non-maximal
atmospheric mixing angle $\theta_{23}$ in the $\nu_\mu$ and
$\bar\nu_\mu$ channels.
The impact of this preference on the determination of $\theta_{23}$ in
our global fit will be discussed in the next section.

\subsection{Global fit results}
\label{sec:results}

Here we summarize the results for the neutrino oscillation parameters
obtained in our present global analysis. For details on the numerical
analysis of all the neutrino samples see
Refs.~\cite{Schwetz:2011qt,Schwetz:2011zk,schwetz:2008er,maltoni:2004ei}
and references therein.


\begin{figure}
  \centering
 \includegraphics[width=0.45\textwidth]{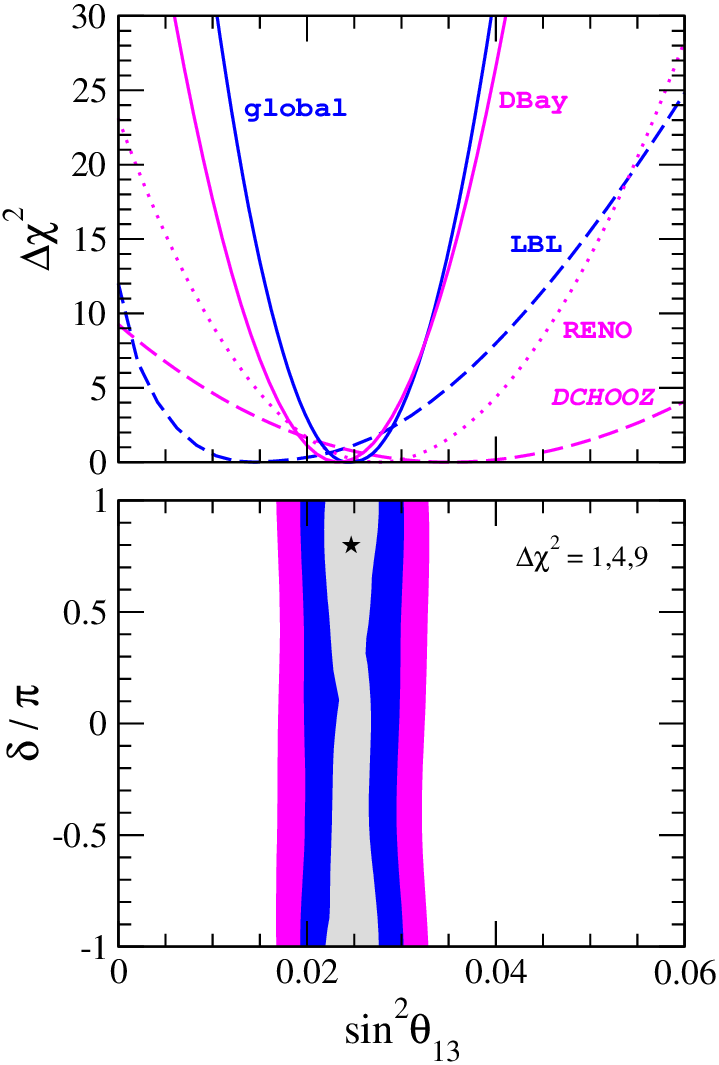}
  \includegraphics[width=0.45\textwidth]{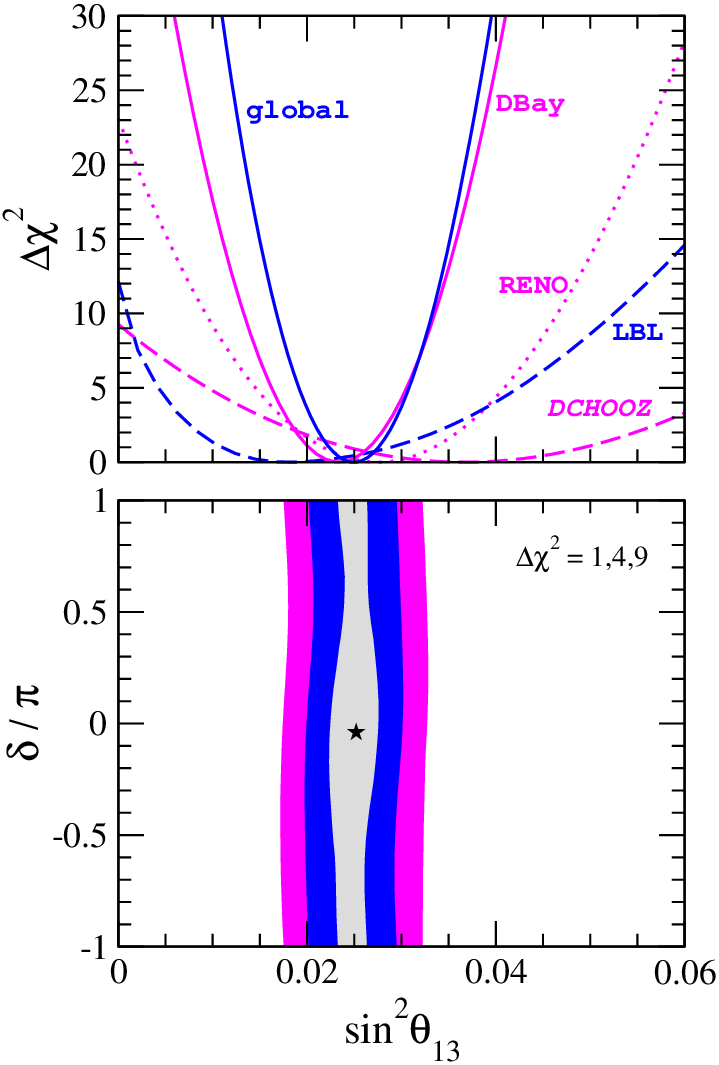}
  \caption{Upper panels: $\Delta\chi^2$ as a function of
    $\sin^2\theta_{13}$ from the analysis of the total event rate in
    Daya Bay (solid magenta/light line), RENO (dotted line) and Double
    Chooz (dashed magenta/light line) as well as from the analysis of
    long-baseline (dashed blue/dark line) and global
    neutrino data (solid blue/dark line). Except for the case of the
    global fit here we have fixed the remaining oscillation parameters
    to their best fit values. Lower panels: contours of
    $\Delta\chi^2=1,4,9$ in the $\sin^2\theta_{13}-\delta$ plane from
    the global fit to the data.  We minimize over all undisplayed
    oscillation parameters. Left (right) panels are for normal
    (inverted) neutrino mass hierarchy.}
\label{fig:t13-delta}
\end{figure}

The results obtained for $\sin^2\theta_{13}$ and $\delta$ are
summarized in Fig.~\ref{fig:t13-delta}.
In the upper panels we show the $\Delta\chi^2$ profile as a function
of $\sin^2\theta_{13}$ for normal (left panel) and inverted (right
panel) neutrino mass hierarchies. The solid blue/dark line corresponds
to the result obtained from the combination of all the data samples
while the others correspond to the individual reactor data samples and
the combination of the long-baseline MINOS and T2K appearance and
disappearance data, as indicated. One sees from the constraints on
$\sin^2\theta_{13}$ coming from the different data samples separately
that, as expected, the global constraint on $\theta_{13}$ is dominated
by the recent Daya Bay measurements.
For both neutrino mass hierarchies we find that the 3$\sigma$
indication for $\theta_{13} > 0$ obtained in our previous
work~\cite{Schwetz:2011zk} due mainly to the first indications
observed by MINOS and T2K is now overwhelmingly confirmed as a result
of the recent reactor data. Thus, in our global fit we obtain a
  $\Delta\chi^2 \sim 104$, resulting in a $10.2\sigma$
 exclusion of $\theta_{13} = 0$ for both mass hierarchies.
In the lower panels of Fig.~\ref{fig:t13-delta} we show the contours
of $\Delta\chi^2=1,4,9$ in the $\sin^2\theta_{13}-\delta$ plane from
the global fit to the neutrino oscillation data.
In this plane we find the following best fit points:  
\begin{eqnarray}
\sin^2\theta_{13} = 0.0246\,,& \, \delta = 0.80\pi & \text{(normal hierarchy),} \\
\sin^2\theta_{13} = 0.0250\,,& \quad\delta = -0.03\pi & \text{(inverted hierarchy).} 
\end{eqnarray}
In our previous analysis~\cite{Schwetz:2011zk} there was a ``preferred
region'' at $\Delta\chi^2 = 1$ for the CP phase $\delta$ for normal
neutrino mass ordering, as a result of the complementarity between
MINOS and T2K appearance data.  One sees that this effect has been
diluted after the combination with the new reactor data, so no
``preferred region'' for the CP phase $\delta$ remains at
$\Delta\chi^2 = 1$~\footnote{Note that, given
      the approximations adopted in the atmospheric neutrino analysis
      in Ref.~\cite{Wendell:2010md}, the sensitivity to the parameter
      $\delta$ in our global fit comes only from long-baseline
      neutrino data.}.
For this reason we marginalize over the CP phase $\delta$ (and all
other oscillation parameters), obtaining for the best fit, one-sigma
errors, and the significance for $\theta_{13} > 0$:
\begin{equation}
\begin{array}{c@{\qquad}l}
\sin^2\theta_{13} = 0.0246^{+0.0029}_{-0.0028}\,,\quad \Delta\chi^2 =103.5 \,(10.2\sigma) & \text{(normal),} \\
\sin^2\theta_{13} = 0.0250^{+0.0026}_{-0.0027}\,,\quad \Delta\chi^2 = 104.7 \,(10.2\sigma) & \text{(inverted).} 
\end{array}
\end{equation}

When compared with the previous analysis in
Refs.~\cite{Schwetz:2011qt,Schwetz:2011zk}, we remark that here we are
not including previous short baseline reactor experiments, which would
lead to a somewhat less significant result for the exclusion of
$\theta_{13} = 0$.

\begin{figure}
  \centering
 \includegraphics[width=0.9\textwidth]{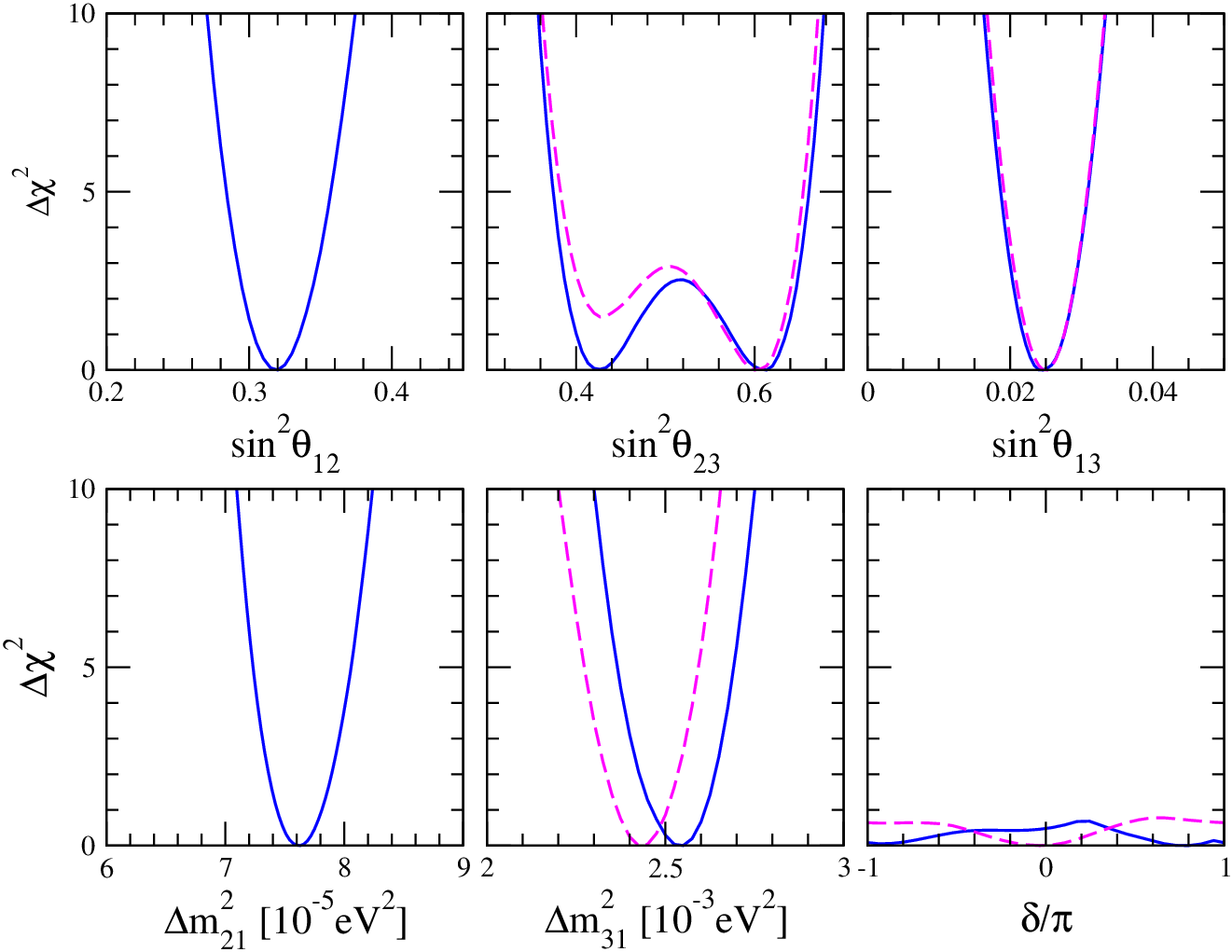}
 \caption{$\Delta\chi^2$ profiles as a function of all the neutrino
   oscillation parameters $\sin^2\theta_{12}$, $\sin^2\theta_{23}$,
   $\sin^2\theta_{13}$, $\Delta m^2_{21}$, $\Delta m^2_{31}$ and
   $\delta$. For the central and right panels the solid lines
   correspond to the case of normal mass hierarchy while the dashed
   lines correspond to the results for the inverted mass hierarchy.}
\label{fig:summary}
\end{figure}

Besides $\theta_{13}$ and $\delta$, from the global analysis of
neutrino data we also recalculate the best fit values and ranges
allowed for all the other neutrino oscillation parameters. Our results
are summarized in Fig.~\ref{fig:summary} and Table~\ref{tab:summary}.
\begin{table}[t]\centering
  \catcode`?=\active \def?{\hphantom{0}}
  \begin{tabular}{|c|c|c|c|c|}
    \hline
    parameter & best fit & $1\sigma$ range& 2$\sigma$ range& 3$\sigma$ range
    \\
    \hline
    $\Delta m^2_{21}\: [10^{-5}\eVq]$
    & 7.62 & 7.43--7.81  & 7.27--8.01 & 7.12--8.20 \\[3mm] 
    $|\Delta m^2_{31}|\: [10^{-3}\eVq]$
    &
    \begin{tabular}{c}
      2.55\\
      2.43
    \end{tabular}
    &
    \begin{tabular}{c}
      $2.46-2.61$\\
      $2.37-2.50$\\
    \end{tabular}
    &
    \begin{tabular}{c}
      $2.38-2.68$\\
      $2.29-2.58$
    \end{tabular}
    &
    \begin{tabular}{c}
      $2.31-2.74$\\
      $2.21-2.64$
    \end{tabular}
    \\[6mm] 
    $\sin^2\theta_{12}$
    & 0.320 & 0.303--0.336 & 0.29--0.35 & 0.27--0.37\\[3mm]  
    $\sin^2\theta_{23}$
    &
    \begin{tabular}{c}
      0.613 (0.427)\footnote{This is a local minimum in the first octant of
        $\theta_{23}$ with $\Delta\chi^2 = 0.02$ with respect to the
        global minimum}\\
      0.600
    \end{tabular}
    &
    \begin{tabular}{c}
      0.400-0.461 \& 0.573--0.635\\
      0.569--0.626\\
    \end{tabular}
    &
    \begin{tabular}{c}
      0.38--0.66\\
      0.39--0.65
    \end{tabular}
    & 
    \begin{tabular}{c}
      0.36--0.68\\ 
      0.37--0.67 
    \end{tabular}
    \\[5mm] 
    $\sin^2\theta_{13}$
    &
    \begin{tabular}{c}
      0.0246\\
      0.0250
    \end{tabular}
    &
    \begin{tabular}{c}
      0.0218--0.0275 \\
      0.0223--0.0276
    \end{tabular}
    &
    \begin{tabular}{c}
      0.019--0.030\\ 
      0.020--0.030     
    \end{tabular}
    &
    0.017--0.033 \\
    $\delta$
   &
   \begin{tabular}{c}
     ?$0.80\pi$\\
     $-0.03\pi$
   \end{tabular}
   &
   $0-2\pi$
 &
   $0-2\pi$
   &
   $0-2\pi$ \\
       \hline
     \end{tabular}
     \caption{ \label{tab:summary} Neutrino oscillation parameters
       summary. For $\Delta m^2_{31}$, $\sin^2\theta_{23}$, $\sin^2\theta_{13}$, 
       and $\delta$ the upper (lower) row corresponds to normal (inverted)
       neutrino mass hierarchy.}
\end{table}

Comparing with our previous results we see that the inclusion of the
new reactor and long-baseline data does not have a strong impact on
the determination of the solar neutrino oscillation parameters, which
are already pretty well determined by solar and KamLAND reactor data.
The differences between the results in Table~\ref{tab:summary} and
those in Table I in \cite{Schwetz:2011zk} are due to the different
treatment of reactor data. Indeed, motivated by the so-called
``reactor antineutrino anomaly''~\cite{Mention:2011rk}, old data from
reactor experiments were included in the analysis in
\cite{Schwetz:2011zk}. The dependence of the determination of solar
neutrino oscillation parameters $\sin^2\theta_{12}$ and $\Delta
m^2_{21}$ upon the details of the reactor data analysis has already
been discussed in detail in Ref.~\cite{Schwetz:2011qt}.

Concerning atmospheric neutrino parameters, the best fit values for
the atmospheric mass splitting parameter $\Delta m^2_{31}$ in
Tab.~\ref{tab:summary} have been shifted to somewhat larger values
compared to our previous results in Ref.~\cite{Schwetz:2011zk}. This
is mainly due to the new MINOS disappearance data in
Ref.~\cite{new-MINOS}, that prefer values for the mass splitting
parameter larger than in their previous data release in
\cite{Adamson:2011ig}.  The precision in the determination of $\Delta
m^2_{31}$ has also been improved thanks to the new long-baseline
neutrino data. Thus, at 3$\sigma$ we find approximately a 8\%
accuracy in the determination of $\Delta m^2_{31}$, while a 12\%
accuracy was obtained in \cite{Schwetz:2011zk} at 3$\sigma$.
For the atmospheric mixing angle we note a slight rejection for
maximal values of $\theta_{23}$. In particular, our global fit shows a
preference for the mixing angle in the second octant. This preference
is very weak for the normal mass hierarchy case, where a local best
fit point at $\sin^2\theta_{23}$ = 0.427 appears with $\Delta\chi^2
\simeq 0.02$, so that a symmetric $\Delta\chi^2$ profile can be seen
at middle-top panel of Fig.~\ref{fig:summary} . For inverted mass
ordering however, the profile is more asymmetric and a local minimum
for $\theta_{23}$ appears in the first octant only at $\Delta\chi^2
\simeq 1.5$.  Maximal mixing, i.e. $\theta_{23} = \pi/4$ is
disfavoured at $\sim$ 90\% C.L.  for both hierarchies.
While the preference for non-maximal values of the atmospheric mixing
angle comes directly from the new MINOS data, the choice of a
particular octant comes from the interplay of long-baseline, reactor
and atmospheric neutrino data, as we will discuss in detail in the
next section.

\section{Discussion}
\label{sec:discussion}

\begin{figure}
  \centering
 \includegraphics[width=0.75\textwidth]{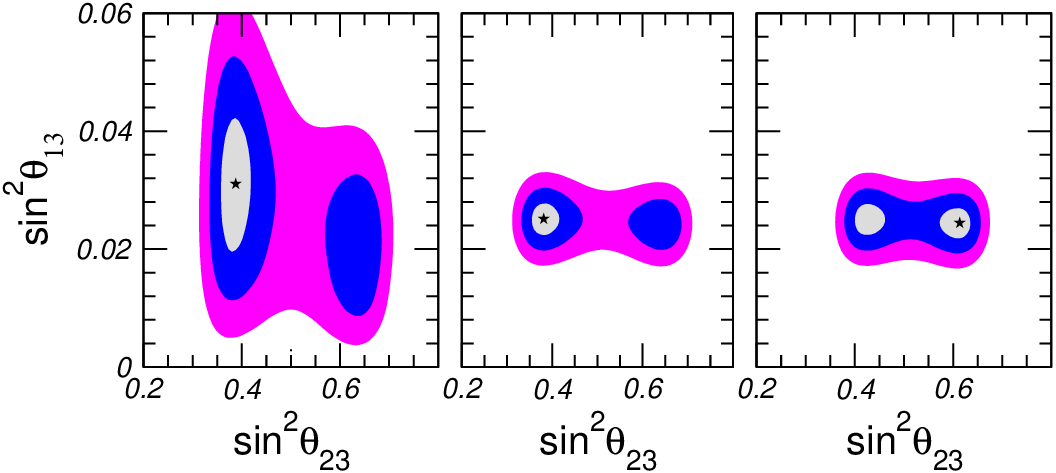}
 \includegraphics[width=0.75\textwidth]{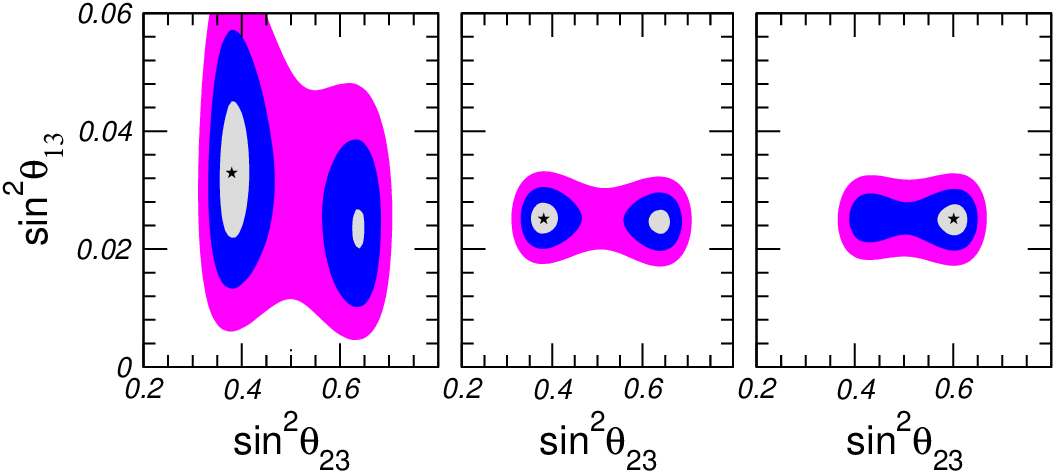}
 \caption{Upper panels: contour regions with $\Delta \chi^2$ = 1, 4, 9
   in the $\sin^2\theta_{23}$ - $\sin^2\theta_{13}$ plane from the
   analysis of long--baseline (MINOS and T2K) + solar + KamLAND data
   (left panel), long-baseline + solar + KamLAND + new Double Chooz,
   Daya Bay and RENO reactor data (middle panel) and the global
   combination (right panel) for normal hierarchy.  Lower panels, same
   but for (inverted) neutrino mass hierarchy.}
\label{fig:t23-disc}
\end{figure}

We now discuss the impact of the new long-baseline data and the
atmospheric neutrino analysis in the determination of the atmospheric
mixing parameter $\theta_{23}$.
As already stated above the new disappearance data from MINOS show a
preference for non-maximal values of $\theta_{23}$.  Due to the
smallness of the associated matter effects in MINOS, these data are
octant-symmetric and therefore say nothing about the octant of the
atmospheric mixing angle $\theta_{23}$.  However, the interplay with
long-baseline neutrino appearance and reactor antineutrino data breaks
the octant-degeneracy, leading to a small preference for values of
$\theta_{23}$ smaller than $\pi$/4.
This is seen in the left panels of Fig.~\ref{fig:t23-disc} where we
have plotted the allowed regions in the $\sin^2\theta_{23}$ -
$\sin^2\theta_{13}$ plane from the combination of long-baseline
(MINOS and T2K) with solar + KamLAND.  These data samples prefer
$\theta_{23}$ values in the first octant and the same holds for the
case when new reactor data are included, see middle panels in
Fig.~\ref{fig:t23-disc}.
Up to this point all statistical analysis are in
agreement~\cite{Fogli:2012ua,nufitcoll}.
 
Nevertheless, when we then include atmospheric data in the global
analysis, differences in the determination of $\theta_{23}$ arise due
to the differences in the analysis of atmospheric neutrino data.
In fact, one sees that the effect of combining with the atmospheric
neutrino data maintains the preference for non-maximal values of
$\theta_{23}$ but leads to a shift towards $\sin^2\theta_{23}$ in the
second octant for both mass orderings, as seen in the right panels in
Fig.~\ref{fig:t23-disc}. In contrast
Refs.~\cite{Fogli:2012ua,nufitcoll} find $\theta_{23}$ in the first
octant for both spectra.  Note however that the preference for a given
octant in our analysis is still rather marginal, and $\theta_{23}$
values in the first octant appear with $\Delta\chi^2$ = 0.02 and 1.5
for normal and inverse mass hierarchy respectively.

As stated in Section~\ref{sec:accel-exper}, for our global fit we use
the official Super-Kamiokande analysis of atmospheric neutrino data in
Ref.~\cite{Wendell:2010md}, performed within the one-mass scale
approximation. This analysis shows a preference for maximal
$\theta_{23}$ mixing, although a small deviation of $\theta_{23}$ to
the second octant is found in the case of inverse mass ordering.
The most recent analysis of the Super-Kamiokande collaboration
presented in Neutrino-2012~\cite{new-SK-atm} and performed with the
inclusion of solar mass splitting corrections is in agreement with
their previous results (obtained without these corrections). In this
case a small preference for $\theta_{23}$ in the first octant for the
normal spectrum, and for second octant for the inverted one is found,
with maximal mixing well inside the one sigma range.
In contrast, the analyses of Refs.~\cite{Fogli:2012ua,nufitcoll},
updated after Neutrino-2012, find a global preference for $\theta_{23}$
in the first octant and exclude maximal mixing at the 2$\sigma$ level
(for normal hierarchy), in qualitative agreement with each other, though
the agreement is not perfect at the quantitative level.
Both of these analyses are at odds with the latest Super-Kamiokande
atmospheric neutrino data analysis in \cite{new-SK-atm}.  At the
moment it is not clear what is the origin of this discrepancy.

The impact of the atmospheric neutrino analysis upon the determination
of $\theta_{23}$ is very visible, therefore in order to get a robust
measurement of the atmospheric mixing angle it is crucial to clarify
the origin of the discrepancies among the various analysis of
atmospheric data.

\section{Conclusion and outlook}
\label{sec:conclusion-outlook}

We have updated the global fit of neutrino oscillation parameters
including the recent measurements of reactor antineutrino
disappearance reported by the Double Chooz, Daya Bay and RENO
experiments, as well as latest MINOS and T2K appearance and
disappearance results, as presented at the Neutrino-2012 conference.
We have found that the preferred global fit value of $\theta_{13}$ is
$\sin^2\theta_{13} = 0.0246 (0.0250)$ for normal (inverted) neutrino mass
hierarchy, while $\sin^2\theta_{13} = 0$ is now excluded at 10.2$\sigma$.
There is reasonable agreement with the results of other
global analyses~\cite{nufitcoll,Fogli:2012ua}, except for the
atmospheric neutrino mixing parameter. We find that the global
analysis pushes the atmospheric mixing angle $\sin^2\theta_{23}$ best
fit value towards the second octant for both neutrino mass orderings.
This hint, however, is still quite marginal and first-octant values of
$\theta_{23}$ are well inside the 1$\sigma$ range for normal hierarchy
and at 1.2$\sigma$ for the inverted spectrum.
Independent phenomenological analyses of atmospheric neutrino data in
Refs.~\cite{Fogli:2012ua,nufitcoll} obtain a preference for mixing
angle in the first octant for both mass hierarchies.  Moreover, the
new official Super-Kamiokande analysis in Ref.~\cite{new-SK-atm} with
full three flavour effects gives a somewhat weaker preference for
non-maximal $\theta_{23}$ mixing, together with a correlation between
the neutrino mass ordering and the preferred octant for $\theta_{23}$.
The origin of this discrepancy which crucially affects the
determination of the atmospheric mixing angle is not yet clear.
The impact of the new reactor and long-baseline accelerator
measurements upon the solar neutrino oscillation parameters is
completely marginal, the results are summarized in
Table~\ref{tab:summary}.

During the summer this year the Daya Bay Collaboration will complete
the designed number of detectors by adding one detector in the far
hall and other one in one of the near halls, re-starting the data
taking after summer with eight neutrino detectors. After 3 years of
operation the uncertainties on $\sin^22\theta_{13}$ will be reduced
from 20\% to 4-5\% \cite{talk:Cao}.
Needless to say that a good determination of a sizeable $\theta_{13}$
value will be a crucial ingredient towards a new era of CP violation
searches in neutrino
oscillations~\cite{nunokawa:2007qh,Bandyopadhyay:2007kx} and will also
help determining the neutrino mass hierarchy.

\vspace*{3mm}

Note added: During the refereeing process of the present work, a new
analysis has appeared in Ref.~\cite{GonzalezGarcia:2012sz}. The
authors obtain a preference for non-maximal values of $\theta_{23}$ at
the 1.7-2$\sigma$ level as a result of the new MINOS disappearance
data. Their new atmospheric neutrino data analysis results in a
reduced sensitivity to the $\theta_{23}$ octant in their global
fit, in better agreement with the analysis by the Super-K
collaboration than their previous results given in \cite{nufitcoll}.

\section*{Acknowledgments}

This work was supported by the Spanish MINECO under grants
FPA2011-22975 and MULTIDARK CSD2009-00064 (Consolider-Ingenio 2010
Programme), by Prometeo/2009/091 (Generalitat Valenciana) and by the EU
ITN UNILHC PITN-GA-2009-237920.  M.T.\ acknowledges financial support
from CSIC under the JAE-Doc programme, co-funded by the European
Social Fund.

\bibliographystyle{h-physrev4}

\end{document}